\documentclass[aps,prb,twocolumn,twoside,notitlepage,nofootinbib,longbibliography]{revtex4-1}
\usepackage{textcomp,hyperref,enumerate, url, cancel}                    			
\usepackage[margin=0.8in]{geometry}
\usepackage{color} 
\usepackage{amsmath,amssymb,amsthm,bbm,bm,nicefrac}   				
\usepackage{graphicx,epsfig}                                                               		
\usepackage{multirow, comment,subfigure}                     
\usepackage[normalem]{ulem}

\newcommand{\ket}[1]{\left|#1\right\rangle}							
\newcommand{\bra}[1]{\left\langle#1\right|}

\newcommand{\abs}[1]{\left\lvert #1\right\rvert}

\newcommand{\be}{\begin{equation}} 							
\newcommand{\ee}{\end{equation}}
\newcommand{\bematrix}{\left(\begin{matrix}}
\newcommand{\ematrix}{\end{matrix}\right)}
\theoremstyle{definition}

\theoremstyle{theorem}

\theoremstyle{lemma}

\theoremstyle{proposition}

\theoremstyle{corollary}

\theoremstyle{observation}

\theoremstyle{remark}

\usepackage{algorithm}
\usepackage{tabularx}
\makeatletter
\def\BState{\State\hskip-\ALG@thistlm}
\makeatother
\usepackage{algpseudocode}
\makeatletter
\algnewcommand{\LineComment}[1]{\Statex \hskip\ALG@thistlm \(\triangleright\) #1}
\makeatother

\makeatletter
\newcommand{\multiline}[1]{%
  \begin{tabularx}{\dimexpr\linewidth-\ALG@thistlm}[t]{@{}X@{}}
    #1
  \end{tabularx}
}
\makeatother

\usepackage{algpseudocode}

\def\one{{\mbox{$1 \hspace{-1.0mm}  {\bf l}$}}}						
\def\C{{\ensuremath{\mathbbm C}}}

\def\ii{\mathrm{i}}

\def\cH{\mathcal H}

\newcommand\defn[1]{\textsl{#1}} 
\newcommand{\rl}[0]{{\rangle\langle}}

\usepackage{nameref}

\begin{document}

\title{How long does it take to implement a projective measurement?}

 \author{Philipp Strasberg$^1$}
 \author{Kavan Modi$^2$}
 \author{Michalis Skotiniotis$^1$}
 \affiliation{$^1$F\'isica Te\`orica: Informaci\'o i Fen\`omens Qu\`antics, Departament de F\'isica, Universitat Aut\`onoma de
 Barcelona, 08193 Bellaterra (Barcelona), Spain}
 \affiliation{$^2$School of Physics and Astronomy, Monash University, Clayton, Victoria 3800, Australia}

\date{\today}

\begin{abstract}
According to the Schr\"odinger equation, a closed quantum system evolves continuously in time. If it is subject 
to a measurement however, its state changes randomly and discontinuously, which is mathematically described 
by the projection postulate. But how long does it take for this discontinuous change to occur? Based on simple 
estimates, whose validity rests solely on the fact that all fundamental forces in nature are finite-ranged, we show 
that the implementation of a quantum measurement requires a minimum time. This time scales proportionally with the 
diameter of the quantum mechanical object, on which the measured observable acts non-trivially, with the proportionality 
constant being around $10^{-5}$~s/m. We confirm our bound by comparison with experimentally reported measurement 
times for different platforms. We give a pedagogical exposition of our argumentation introducing along the way 
modern concepts such as ancilla-based measurements, the quantum speed limit, and Lieb-Robinson velocity bounds.
\end{abstract}
\maketitle

\section{Introduction}
The process of measurement in quantum theory is radically different as compared to any other physical 
theory.\cite{WignerAJP1963, BellPW1990} Upon observing a quantum mechanical system its state randomly and 
\emph{discontinuously} changes into an eigenstate of the corresponding observable. It is this unpredictable 
change that most students have trouble coming to terms with as there is no analogue in classical mechanics. 
Indeed, the measurement or projection postulate remains a highly active area of research\cite{PokornyEtAlPRL2020} 
and is at the heart of interpretational issues concerning quantum theory.\cite{CabelloInBook2017} For instance, 
within the commonly taught Copenhagen interpretation of quantum mechanics, this change is epitomized by the 
``collapse'' of the wave function. 

Since quantum mechanics was conceived for the microworld, a naturally arising question is which postulates can 
be carried over to the macroworld. Here, we explain that the time to \emph{implement} a measurement is necessarily 
finite. For macroscopic observables, this time can be much longer than the time scales of intrinsic quantum mechanical 
processes of the object itself, posing severe practical challenges for the application of the projection postulate. 
As we will explain, the minimum measurement time for a quantum mechanical observable that acts non-trivially on 
a physical system of size $d$ is 
\begin{equation}\label{eq Strasi Skoti Modi time}
 t_\text{min}\geq \frac{d}{v} \qquad \mbox{with} \qquad v = 10^{5}~\frac{\text{m}}{\text{s}}\; ,
\end{equation}
in good agreement with state-of-the-art experiments. Thus, the larger the measured object the longer it takes 
to measure it, highlighting that the discontinuity in the measurement postulate has to be taken \emph{cum grano 
salis} (with a grain of salt). We remark that Eq.~(\ref{eq Strasi Skoti Modi time}) does not hold for photons 
but for systems composed of electrons, atoms or molecules. Importantly, our conclusions hold independently of the 
reader's preferred interpretation of quantum theory. 

We derive Eq.~(\ref{eq Strasi Skoti Modi time}) armed with nothing more than quantum theory itself and the 
fact that all forces in nature are finite-ranged. Our line of reasoning makes use of three important tools which 
are absent from most standard quantum mechanics curricula; \emph{ancilla-assisted measurements}, which will help us 
to mathematically formulate the process of measurement in terms of the more familiar unitary evolution postulate of 
quantum theory, \emph{quantum speed limits} which tell us that evolution in a Hilbert space takes finite time, and 
Lieb-Robinson bounds which limit the speed of propagation of information in space. We start with a pedagogical 
exposition of these techniques in Sec.~\ref{sec:QTvsQP}, which find applications in a wide range of problems, 
such as quantum sensing, metrology, and quantum thermodynamics, and form an active area of current research. 
We then employ these results to establish a lower bound on the time it takes to measure a macroscopic object of size 
$d$ in Sec.~\ref{sec time}. There we also provide a back-of-the-envelop comparison of our bound with experimentally 
reported measurement times in quantum systems. We summarise and conclude in Sec.~\ref{sec discussion}. 
 
\section{Review of pertinent concepts}
\label{sec:QTvsQP}

We start in Sec.~\ref{subsec maths} with a brief review of the mathematical postulates of quantum \emph{theory} 
followed in Sec.~\ref{sec:ancilla-based} with a discussion about an important mathematical tool referred to as 
ancilla-assisted measurements. Whereas these first two sections are free from any physical considerations, we introduce 
physical constraints in the remaining three sections. These include energy constraints on the minimal evolution time of 
a quantum system (quantum speed limits, Sec.~\ref{subsec qsls}), constraints on the form of the admissible 
Hamiltonians, which have to be local due to the finite range of all fundamental forces 
(Sec.~\ref{sec local interactions}), and constraints on the speed at which information can propagate in locally 
interacting objects (Lieb-Robinson bounds, Sec.~\ref{sec objects and observables}). We call the framework that results 
from supplementing the mathematical postulates of quantum theory with physical constraints quantum \emph{physics}. 

\subsection{The mathematical postulates of quantum theory}
\label{subsec maths}

For our purposes it suffices to consider the mathematical formalism of quantum theory as it applies to closed systems.  
The state of such a system is described by a vector $\ket{\psi}\in\cH$ in a Hilbert space $\cH$ or, more 
generally, by a positive-definite operator of unit trace, $\rho$, the {\it density matrix}.  The closed-system evolution 
of such a system is described by a unitary operator, $U_t: \cH\to\cH$, with $U_tU^\dagger_t = U^\dagger_t U_t =\one$, 
the identity matrix, and $t$ the time. Measurements of an observable $R = \sum^{n-1}_{r=0} \lambda_r \Pi(r)$ are 
described by sets of projection operators $\{\Pi(r)\}$, with $r$ labeling the possible measurement outcomes (assumed to 
be discrete for simplicity), satisfying the conditions 
$\Pi(r)\Pi(s)= \delta_{r,s} \Pi(r)$ ($\delta_{r,s}$ the usual Kronecker delta) and $\sum_r \Pi(r)=\one$. 
The probability of obtaining a given outcome, $r$, is given by Born's rule, $p(r) = \mbox{tr}\{\Pi(r)\rho\}$. Upon 
obtaining the outcome $r$, the {\it post-measurement} state of the system changes to 
\begin{equation}\label{eq collapse}
 \rho(r) = \frac{\Pi(r)\ \rho \ \Pi(r)}{p(r)}.
\end{equation}
Finally, the state space of a composite system, made out of $N$ distinguishable constituents whose Hilbert spaces 
are $\cH_i, \, i\in(1,\ldots,N)$, is given by the tensor product 
$\cH=\cH_1\otimes\cH_2\otimes\ldots\otimes\cH_N:=\bigotimes_{i=1}^N\,\cH_i$.

We assume that the time evolution of all objects in our discussion, including macroscopic ones, are described 
by Schr\"odinger's equation.  Consequently,  we also model the interaction between an isolated quantum system and any object which one could potentially regard as measurement apparatus
(including a human brain) by a unitary operator. This unitary entangles the system with the apparatus without 
either experiencing wavefunction collapse as per Everett's ``many-worlds'' interpretation.\cite{EverettRMP1957} 
We now review these \defn{ancilla-based} measurement schemes which form part of the standard toolkit of quantum 
measurement theory.\cite{HolevoBook2001b, DAlessandroBook2007, WisemanMilburnBook2010, JacobsBook2014}
 
\subsection{Ancilla-based measurements}
\label{sec:ancilla-based}

\begin{figure}[t!]
	\includegraphics[keepaspectratio, width=8cm]{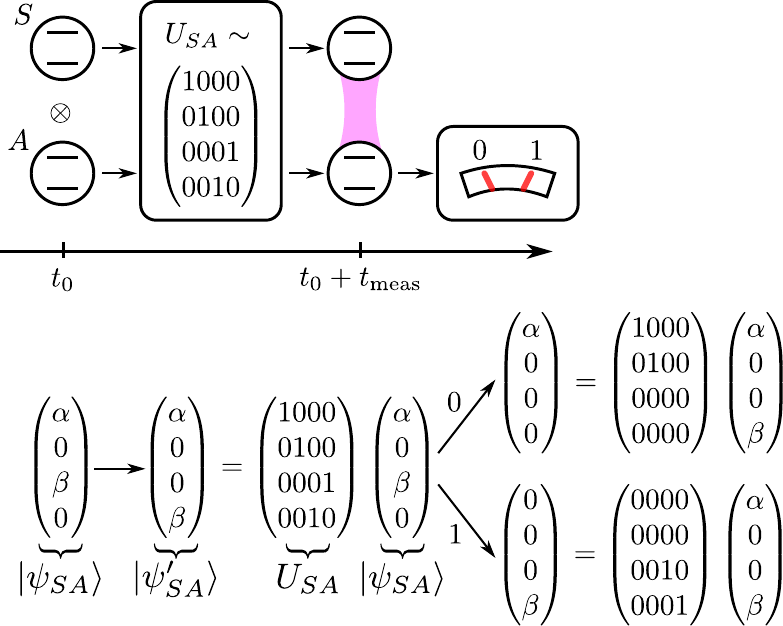}
	\caption{An ancilla-based strategy for implementing the projective measurement 
	$\{\Pi_S(0)=|0\rl0|_S,\, \Pi_S(1)=|1\rl1|_S\}$ (equivalently implementing a measurement of the 
	observable, $\sigma^{(z)} =|1\rl1| - |0\rl0|$).  The system is a two-level system initially in state
	$\ket{\psi}=\alpha\ket{0}+\beta\ket{1}$, where $\alpha,\beta\in\C$ with $\abs{\alpha}^2+\abs{\beta}^2=1$. The 
	ancilla is also a two-level system initially prepared in state $\ket{0}$. They interact via the Hamiltonian 
	$H_{SA} = \frac{\hbar g}{2}[|0\rl0|_S\otimes \one_A + |1\rl1|_S\otimes(\one+\sigma^{(x)})_A]$ 
	(with $\sigma^{(x)} = |0\rl1|+|1\rl0|$) and evolve unitarily for a time $t=\pi/g$. Up to a global phase this 
	implements the transformation $U_{SA}\ket{\psi}_{SA}=\alpha\ket{00}_{SA}+\beta\ket{11}_{SA}$. Measuring the ancilla 
	with the projective measurement $\{\Pi_A(0)=|0\rl0|_A,\, \Pi_A(1)=|1\rl1|_A\}$, we obtain $r=0$ with 
	probability $p(0)=\lvert\alpha\rvert^2$ and $r=1$ with probability $p(1)=\lvert\beta\rvert^2$. The corresponding 
	post-measurement states of the system are $\ket{0}$ and $\ket{1}$, respectively. Vectors and matrices shown in the 
	figure are with respect to the ordered basis 
	$\{|00\rangle_{SA},|01\rangle_{SA},|10\rangle_{SA},|11\rangle_{SA}\}$. }
	\label{fig:zmeas}  
\end{figure}

To explain the idea of ancilla-based measurements, which goes back to von Neumann, it is useful to look at an 
example first (see also Fig.~\ref{fig:zmeas}). Consider a two-level system in the state 
$|\psi\rangle_S = \alpha|0\rangle_S + \beta|1\rangle_S$ with 
$|\alpha|^2 + |\beta|^2 = 1$ and suppose we want to perform a measurement in the basis $\{|0\rangle,|1\rangle\}$. 
To this end, we let it interact with an external \emph{probe} or \emph{detector}, called an \emph{ancilla} in the 
following, which is itself a quantum system. Quite intuitively, the measurement of a 
two-level system requires not more than a two-level system, which we assume to be prepared in the state 
$|0\rangle_A$. Thus, the initial system-ancilla state is $|\Psi(0)\rangle_{SA} = |\psi\rangle_S\otimes|0\rangle_A$. 
Now, we let the system and ancilla interact in such a way that the unitary after time $t$ is 
\begin{equation}\label{eq system ancilla unitary TLS}
 U_{SA}(t) = |00\rl00| + |01\rl01| + |10\rl11| + |11\rl10|,
\end{equation}
where we used the shorthand notation $|00\rangle \equiv |0\rangle_S\otimes|0\rangle_A$, etc. What is the state after 
the interaction? A quick calculation reveals 
\begin{equation}
 |\Psi(t)\rangle_{SA} = U_{SA}(t)|\Psi(0)\rangle = \alpha|00\rangle + \beta |11\rangle,
\end{equation}
which is a perfectly correlated state: the system is in state $|0\rangle$ ($|1\rangle$) if and only if the 
ancilla is in state $|0\rangle$ ($|1\rangle$). Measuring the state of the ancilla therefore reveals the state of the 
system and implements the desired projective measurement. It is interesting to note that the reduced state of the 
system (as well as of the ancilla) is \emph{mixed}, 
\begin{equation}
 \mbox{tr}_A\{|\Psi(t)\rl\Psi(t)|_{SA}\} = |\alpha|^2 |0\rl0|_S + |\beta|^2 |1\rl1|_S,
\end{equation}
which is a consequence of the quantum entanglement between system and ancilla. Another quick calculation reveals 
that this state is identical to 
\begin{equation}
 |0\rl0|\psi\rl\psi|0\rl0|_S + |1\rl1|\psi\rl\psi|1\rl1|_S,
\end{equation}
which corresponds to the average effect of a measurement in the basis $\{|0\rangle,|1\rangle\}$. It is therefore also 
called the \emph{unconditional} measurement state because it is not yet conditioned on receiving the outcome $0$ or $1$. 
In fact, it is impossible to implement Eq.~(\ref{eq collapse}) in a unitary way, which in the present case means 
a transformation of the form $|0\rl0|\psi\rl\psi|0\rl0|_S/|\alpha|^2$ if $r=0$ or 
$|1\rl1|\psi\rl\psi|1\rl1|_S/|\beta|^2$ if $r=1$. 
The reason is that Eq.~\eqref{eq collapse} cannot follow from unitary time-evolution alone as it is a non-linear 
transformation with respect to $\rho$ (recall that the probability $p(r)$ depends on $\rho$ too). 

The main question we are asking in this manuscript is: What is the minimum measurement time $t$ required to 
implement the unitary in Eq.~(\ref{eq system ancilla unitary TLS}), i.e., what is the minimum time required to 
sufficiently correlate a system with an external detection apparatus or ancilla? We will address this question in 
particular for macroscopic systems by using physically motivated requirements. Before we come to them, 
we show that the approach above is general and can be used for any system and measurement. 

We consider an arbitrary system state $\rho_S$ and an arbitrary system observable $R_S$ with $n$ associated 
projectors $\Pi_S(r)$. As above, we assume the ancilla is prepared in the state $|0\rl0|_A$ such that the initial 
system-ancilla state reads $\rho_{SA}(0) = \rho_S\otimes|0\rl0|_A$. 
Next, 
we assume that the experimentalist can engineer an interaction between $S$ and  $A$ resulting in the unitary operator 
\begin{equation}\label{eq system ancilla unitary}
\fontdimen16\textfont2=3pt
U_{SA}(t) =  \sum_{r=0}^{n-1}\, \Pi_S(r)\otimes\sum_{x=0}^{n-1}|x+r\rl x|_A.
\end{equation}
Here, $\{\ket{x}_A\!,\; x\in(0,\ldots n-1)\}$ denotes an arbitrary basis in the ancilla Hilbert space and we interpret 
$x+r$ modulo $n-1$ whenever $x+r>n-1$. It is easy to check that $U_{SA}(t)$ is unitary. 
Furthermore, a straightforward calculation reveals that 
\begin{equation}
 U_{SA}(t)\rho_{SA}(0)U_{SA}^\dagger(t) = \sum_{r,r'} \Pi_S(r)\rho_S\Pi_S(t')\otimes|r\rl r'|_A.
\end{equation}
In particular, the reduced system state becomes 
\begin{equation}
 \rho_S(t) = \sum_r \Pi_S(r)\rho_S \Pi_S(r),
 \label{eq: postmeasurestate}
\end{equation}
which describes the unconditional measurement state as explained above. It is easy to check that this corresponds 
to the average of Eq.~(\ref{eq collapse}): $\rho_S(t) = \sum_r p(r)\rho_S(r)$. 

In reality, the general description above is minimal but somehwat simplistic. However, as we will see, we will 
not put any restrictions on the ancilla in the following. In our view, 
the use of an ancilla corresponds to a minimal theoretical description of the process, which contains all essential 
features to understand it. In present-day experiments, these abstract ancillae can indeed be realized and controlled 
using either physically distinct quantum systems (electrons, photons, {\it etc}.) or even additional degrees of freedom 
of the same physical system, such as the motional degrees of freedom of the quantum system, or higher electronic 
(energy) eigenstates. Whatever the physical realization of the ancilla is, our conclusions will not rely on it. 
	
Before we proceed, we remark that measurements in quantum mechanics are not always described by the projection 
postulate, Eqs.~(\ref{eq collapse}) or~(\ref{eq: postmeasurestate}). An example is the \emph{detection} of particles or 
photons. This measurement certainly reveals information about the state of the system \emph{before} the measurement, 
but afterwards the system no longer `exists' in the sense considered here. Mathematically, it turns out that such a 
state change can be captured by generalizing Eq.~(\ref{eq: postmeasurestate}) to 
$\rho' = \sum_r K_r\rho K_r^\dagger$.~\cite{HolevoBook2001b, DAlessandroBook2007, WisemanMilburnBook2010, 
JacobsBook2014} Here, the operators $K_r$ only need to satisfy the completeness relation 
$\sum_r K_r^\dagger K_r = \one$, but otherwise they are arbitrary (in particular, they do not need to be projectors). 
For simplicity in the presentation, we decide to focus only on Eq.~(\ref{eq: postmeasurestate}) here. However, the 
treatment of the general case parallels our argumentation because it turns out that the map 
$\rho' = \sum_r K_r\rho K_r^\dagger$ can be also implemented by a suitable unitary evolution in a joint 
system-ancilla space. 
 	
\subsection{Quantum speed limits}
\label{subsec qsls}

So far, our exposition was based purely on mathematical considerations. Quantum physics starts to emerge by 
a set of \defn{correspondence rules} that assign dynamical variables, such as position, momentum, {\it etc.}, 
to Hermitian (or self-adjoint) operators.\cite{BallentineBook1998} 

For instance, unitary time evolution can always be written as (which formally follows from Stone's theorem)
\begin{equation}
U_t= \exp\left(-\imath \frac{H}{\hbar} t\right),
\label{eq:expmap}
\end{equation}
where $H$ is some Hermitian operator and $\hbar$ the reduced Planck constant. In quantum theory, any Hermitian 
operator can generate unitary evolution. In quantum physics, on the other hand, unitary evolution is determined by 
the systems Hamiltonian through Schr\"odingers's equation and Eq.~(\ref{eq:expmap}) holds for time-independent 
Hamiltonians only. The crucial distinction between quantum theory and quantum 
physics is that not every Hermitian operator corresponds to a physical Hamiltonian. We shall define precisely what we 
mean by a physical Hamiltonian shortly.  Presently, we are interested in how fast we can implement unitary evolution 
given a Hermitian operator $H$.  
 
The \emph{quantum speed limit} quantifies the minimum time required for a system, initially in state $\ket{\psi}$, 
to evolve to an orthogonal state, $\ket{\psi_\perp}$, under the unitary evolution of 
Eq.~\eqref{eq:expmap}.\cite{MandelstamTammJP1945, MargolusLevitinPD1998} Given a Hermitian operator $H$, the quantum 
speed limit provides a lower bound to the evolution time as  
\be
t\geq \tau_{\text{qsl}}=\max\left\lbrace\frac{\hbar\pi}{2\Delta E},\,\frac{\hbar\pi}{2(\langle H\rangle-E_0)}\right\rbrace.
\label{eq QSL}
\ee
Here, $\Delta E=\sqrt{\langle H^2\rangle-\langle H\rangle^2}$ is the standard deviation of the energy with 
expectation value $\langle H\rangle=\bra{\psi} H\ket{\psi}$, and $E_0$ is the energy of the ground state. The 
bound $\hbar\pi/2\Delta E$ is due to Mandelstam and Tamm,\cite{MandelstamTammJP1945} whereas the bound 
$\hbar\pi/2(\langle H\rangle-E_0)$ is due to Margolus and Levitin.\cite{MargolusLevitinPD1998}

It is worth pausing for a moment to reflect on the motivation behind the Mandelstam and Tamm bound discovered in 1945. 
Shortly after Heisenberg postulated his famous uncertainty principle of position and momentum, an analogous time-energy 
uncertainty relations was postulated based on dimensional arguments. While the position-momentum uncertainty relation 
was formally derived, the time-energy uncertainty relation was difficult to derive as time is not an observable. 
Mandelstam and Tamm resolved this issue with their bound giving the time-energy uncertainty relation firm mathematical 
footing. This simple relation between time and energy is now an integral part of a quantum theorist's 
toolkit.\cite{DeffnerCampbellJPA2017}

By way of example, let us compute the quantum speed limit for any ancilla-based measurement of a binary observable, 
$R_S$, on the system. In this case it suffices to consider a two-dimensional ancilla for its implementation. Denote by 
$\{\ket{0}_A\!,\,\ket{1}_A\!\}$ the orthonormal basis states of the ancilla and assume it is initially prepared in the 
state $\ket{0}_A$.  The unitary of Eq.~\eqref{eq system ancilla unitary} can be implemented by the following 
system-ancilla Hamiltonian (see also Fig.~\ref{fig:zmeas}) 
\begin{equation}\label{eq Hamiltonian binary system ancilla}
 H_{SA} =\frac{\hbar g}{2}\left[|0\rl0|_S\otimes \one_A + |1\rl1|_S\otimes (\one+\sigma^{(x)})_A\right],
\end{equation}
with $\sigma^{(x)}$, $\sigma^{(y)}$ and $\sigma^{(z)}$ the standard Pauli matrices. 
If $H_{SA}$ is ``switched on'' for a time $t=\pi/g$ we obtain 
$U_{SA} = |0\rl0|_S\otimes \one_A + |1\rl1|_S\otimes\sigma_A^x$ as desired. 

Now let us determine the minimum time according to the quantum speed limit. Note that we have $E_0=0$ for the 
Hamiltonian in Eq.~\eqref{eq Hamiltonian binary system ancilla}. If the system-plus-ancilla are originally 
in the state $\ket{00}_{SA}$, then $U_{SA}\ket{00}_{SA}=\ket{00}_{SA}$ and there is no quantum speed limit to apply. 
However, if our system-plus-ancilla state is $\ket{10}_{SA}$, then we find that both bounds in 
Eq.~(\ref{eq QSL}) coincide, giving rise to 
\begin{equation}\label{eq QSL example}
 \tau_\text{qsl}=\frac{\hbar\pi}{2\Delta E_{SA}}=\frac{\hbar\pi}{2\bra{10} H_{SA}\ket{10}}=\frac{\pi}{g}.
\end{equation}
Thus, the interaction Hamiltonian of Eq.~\eqref{eq Hamiltonian binary system ancilla} is the quickest way of implementing the ancilla-based measurement; any other interaction Hamiltonian would require equal or longer time.

\subsection{Locality of interactions}
\label{sec local interactions}

The fact that not every Hermitian operator corresponds to a physical Hamiltonian arises from the fact that physical systems are 
subject to further constraints. For instance, in closed systems energy, momentum, {\it etc.}, are conserved quantities.  
Moreover, for macroscopic physical systems, composed out of a large number of particles with 
non-zero rest mass (such as electrons, nuclei, atoms, or molecules), the dominant interactions are mediated via the 
Coulomb force. The latter gives rise to a set of physical Hamiltonians which describe locally-interacting systems as we now 
explain.

\begin{figure}
 \centering\includegraphics[width=0.25\textwidth,clip=true]{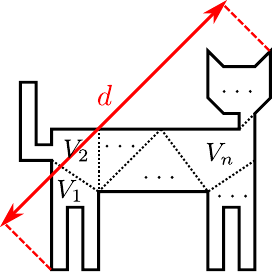}
 \caption{A macroscopic object of (rough) diameter $d$ split into a set of non-overlapping volumes $V_i$. }
  \label{fig cat} 
\end{figure}

Let $H_S$ be the Hamiltonian associated to a physical system occupying a space-like volume, $V$, of diameter $d$ 
(following the spherical-cow tradition in theoretical physics we characterize the size of the object by its diameter 
only). To explain the notion of \defn{interacting} and \emph{local} we split $V$ into an arbitrary set $\{V_i\}$ of 
non-overlapping volumes $V_i$ (i.e., $V_i\cap V_j = \emptyset$ if $i\neq j$, see Fig.~\ref{fig cat}). To each such 
volume we assign a Hilbert space, $\cH_i$, such that the Hilbert space of the entire system is 
$\cH_S = \bigotimes_i \cH_i$. We can now split the Hamiltonian into a sum involving a local and interacting part, 
\begin{equation}
 H_S = H_\text{loc} + H_\text{int},
 \label{eq localH}
\end{equation}
where $H_\text{loc} = \sum_i H_i$ contains all the local Hamiltonians---$H_i$ acting non-trivially only on $\cH_i$---and 
$H_\text{int} = \sum_{i\neq j} J_{ij}$ describes the pair-wise interactions between volumes $V_i$ and $V_j$. The assumption
of pair-wise interactions is ultimately justified by the fact that the Coulomb force is a two-body interaction. We define a 
\emph{locally interacting} physical system if the following two conditions are satisfied:

\begin{enumerate}[(i)]
\item Each $V_i$ interacts at least with one other $V_j$.  Here, by interaction we mean that  $[H_i,J_{ij}] \neq 0$.
\item  For all $i,j$ the strength of the interaction is inversely proportional to the distance, $d_{ij}$, between the two 
locations. If $V_i$ and $V_j$ have themselves a non-negligible diameter, $d_{ij}$ can be defined as an average or minimal 
distance. 
\end{enumerate}

The first condition allows the transfer of energy and information through the physical system, which we assume to avoid 
trivial situations (e.g., objects composed out of non-interacting parts). The second condition arises from the fact 
that all fundamental forces in nature are finite-ranged and decrease with increasing distance. Together, 
conditions (i) and (ii) above set the constraints on the set of physically admissible Hamiltonians 
that we consider here. 

\begin{figure}
 \centering\includegraphics[width=0.42\textwidth,clip=true]{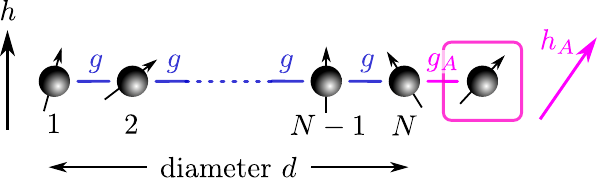}
 \caption{Sketch of a chain of $N$ interacting spins coupled to an ancilla spin (spin in the pink box on the right). 
 The spins in the chain interact via a nearest-neighbour interaction of coupling strength, $g$, and are influenced by 
 an external magnetic field, $h$. The diameter, $d$, of the object is indicated as the length of the chain. To 
 implement the ancilla-based measurement protocol of Sec.~\ref{sec:ancilla-based}, we imagine an external ancilla spin 
 put into contact with the chain. The local magnetic field $h_A$, as well as the local coupling $g_A$ to the $N$th spin 
 of the chain can be externally controlled. }
  \label{fig Ising} 
\end{figure}

To give a specific example, we present a simple model of such a physical system made out of interacting spins on a 
one-dimensional lattice as shown in Fig.~\ref{fig Ising}. The simplest interaction one can conceive of is that of 
nearest-neighbour interactions: each spin interacts only with the spins adjacent to it. Whilst being a strong 
simplification, we note that models restricted to nearest-neighbour interactions can qualitatively 
describe many phenomena in condensed matter physics~\cite{FuldeBook1995, BruusFlensbergBook2004, SachdevBook2007}. The reason 
is that nearest-neighbour interactions often play the most dominant role. In Fig.~\ref{fig Ising}, for example, we 
depict the Ising model whose Hamiltonian
\begin{equation}\label{eq Ising}
 H_\text{Ising} = h\sum_{i=1}^N \sigma_i^{(z)} + g\sum_{i=1}^{N-1} \sigma_i^{(x)}\sigma_{i+1}^{(x)}
\end{equation}
describes $N$ interacting spins in an external magnetic field, $h$, and with spin-spin interaction strength, $g$. 
While it is good to keep the example of the 1D Ising model in mind for illustrative purposes, we remark that 
our conclusions below are general and not restricted to these kind of models. 

\subsection{Finite size systems and the Lieb-Robinson bound}
\label{sec objects and observables}

The objects we are here interested in are composed out of a very large number of locally interacting particles 
with non-zero rest mass (this includes, e.g., electrons, nuclei, atoms, or molecules, but excludes the 
treatment of, e.g., noninteracting photons). To define a \emph{macroscopic} object we estimate the size of each 
constituent by invoking the equipartition theorem. The latter states that the kinetic energy of a single particle at 
temperature $T$ is of the order $mv^2 \approx k_B T$, where $m$ is the mass of the particle, $v$ its velocity, and 
$k_B$ is Boltzmann's constant.  By identifying the particle's momentum via $mv = h/\lambda_\text{th}$ with 
$h = 2\pi\hbar$, we obtain the \emph{thermal de Broglie wavelength} $\lambda_\text{th} = h/\sqrt{mk_B T}$ of a 
single particle. For example, a single hydrogen atom at room temperature has $\lambda_\text{th}\approx 100$~nm. 
According to this reasoning, a physical system is macroscopic if $d\gg\lambda_\text{th}$. 
This crude estimate is sufficient for our purposes, for a more detailed discussion of ``macroscopicity'' 
see Ref.~\cite{JaegerAJP2014}. 

Moreover, we are interested in observing \emph{global} properties of such systems.  For example, in Schr\"odinger's cat 
\emph{gedanken} experiment  we measure the whole cat,  not just its tail, in order to determine its state of well-being.  
The corresponding elements of such global dynamical variables correspond to observables, $R$, that are highly non-local, 
i.e., they act non-trivially on the entire macroscopic object. 

Following Sec.~\ref{sec:ancilla-based}, the goal is now to engineer a suitable interaction between the system and a 
measuring device---the ancilla---such that the unitary of Eq.~\eqref{eq system ancilla unitary} ensues. After suitably 
choosing the initial ancilla state, this gives rise to the post measurement state of Eq.~\eqref{eq: postmeasurestate} as 
desired. Recalling  Eq.~\eqref{eq Hamiltonian binary system ancilla}, the most straightforward way seems to be a 
system-ancilla Hamiltonian of the form 
\begin{equation}\label{eq SA general interaction}
 H_{SA} = \sum_r \,\Pi_S(r) \otimes H_A(r)
\end{equation}
with suitably chosen ancilla Hamiltonians $\{H_A(r)\}$. However,  for highly non-local observables
this implies that the projectors $\Pi_S(r)$ act non-trivially on a large part of the system. It follows that
the Hamiltonian of Eq.~\eqref{eq SA general interaction} is unphysical in that it does not belong to the class of 
Hermitian operators corresponding to locally-interacting physical systems. 

Realistic Hamiltonians have to take into account that the measuring device can only interact locally with the system. 
Returning to the example of the Ising model from Eq.~\eqref{eq Ising}, a more physical system-ancilla interaction 
Hamiltonian is
\begin{equation}\label{eq system ancilla Ising}
 H_{SA} = H_\text{Ising} + h_A\sigma^{(z)}_A+ g_A\sigma^{(x)}_N\sigma^{(x)}_A,
\end{equation}
where we assumed---for illustrative purposes---that the ancilla is represented by a single spin coupled to the $N$th 
spin of the Ising chain from Eq.~(\ref{eq Ising}) (see Fig.~\ref{fig Ising}). 
 
One may wonder whether there is actually any chance at all to implement the unitary of 
Eq.~\eqref{eq system ancilla unitary} based on the limited amount of control offered by 
Eq.~\eqref{eq system ancilla Ising}.  Insights into this question are offered by Zassenhaus' formula (see 
Appendix~\ref{sec appendix Zassenhaus}), a special case of the more general \emph{Baker-Campbell-Hausdorff 
formula}. Despite being a challenging theoretical and experimental task, we assume for the remainder of this work 
that local interactions are sufficient to generate the required unitary in Eq.~\eqref{eq system ancilla unitary}. 

As locally-interacting Hamiltonians satisfy conditions (i) and (ii), information cannot travel arbitrarily fast through 
the physical system, even if we neglect relativistic considerations. A local perturbation needs time before it can 
influence a different region in space. This idea can be formalized as follows. Let $A_i, A_j$ be two observables 
acting on the Hilbert spaces $\cH_i,\,\cH_j$ corresponding to two particles a distance $d_{ij}$ apart. If the 
interaction between sites $i$ and $j$ is finite ranged, then the Lieb-Robinson bound~\cite{LiebRobinsonCMP1972} states
\be \label{eq Lieb Robinson}
\|[A_j(t), A_i]\|\leq C \, \exp [ -a (v_I\, t-d_{ij})],
\ee
where $A_i(t)=e^{\ii t H/\hbar}\, A_i\, e^{-\ii t H/\hbar}$ with $H$ the locally interacting Hamiltonian and 
$a\ge0$ and $C\ge0$ are suitable constants. Here, $\|\cdot\|$ is a suitable operator norm measuring the 
``size'' or ``significance'' of the term on the left hand side of Eq.~(\ref{eq Lieb Robinson}).  Furthermore, 
$v_I\ge0$ is known as the \emph{Lieb-Robinson velocity}, where we use the subscript $I$ to emphasize that this 
determines the speed by which \emph{information} can travel through the system. 

\begin{figure}
 \centering\includegraphics[width=0.46\textwidth,clip=true]{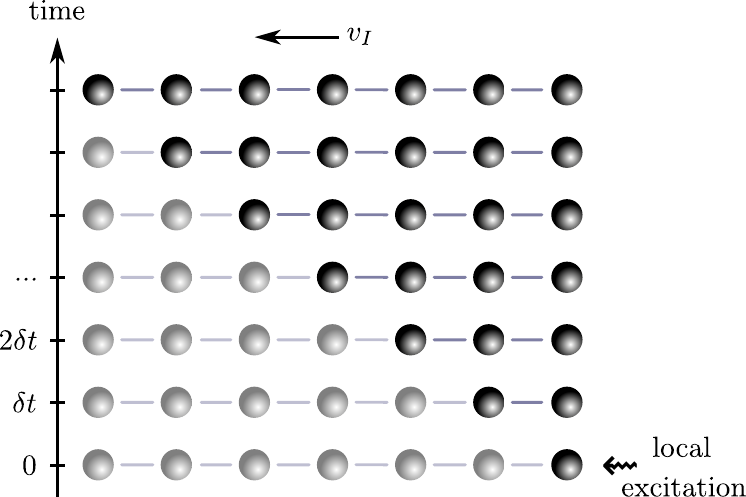}
 \caption{Illustration of the Lieb-Robinson bound. Consider the Ising model from Eq.~\eqref{eq Ising} prepared initially in its ground state (shaded balls). Suppose now that we locally perturb one end of the chain (dark ball and wiggly arrow). How long does it take until the other end of the chain ``feels'' an influence of this perturbation? Computing this time exactly is challenging, but it is easy to understand from the structure of 
 Eq.~\eqref{eq Ising} that this must take a \emph{finite} time. For this purpose assume that we discretize time into 
 small steps $\delta t$ such that the unitary time evolution operator can be approximated as 
 $e^{-iH_\text{Ising}\delta t/\hbar} \approx I - iH_\text{Ising}\delta t/\hbar$. Now, recall that $H_\text{Ising}$ is local and contains only nearest-neighbor interactions. Thus, in the first time step the excited spin on the right of the chain can only influence its nearest  neighbor to the left (now also sketched as a dark ball). In the next time step, this neighbor can again influence only its nearest left neighbor, and so on and so forth until the excitation reaches the leftmost spin. In our example the time taken is $t = 6\delta t$. If one takes the ratio of the 
 distance traveled by the excitation, which here equals the diameter of the object, and this time one obtains the 
 finite velocity $v_I$.}
  \label{fig Lieb Robinson} 
\end{figure}

In our case, $H$ includes the system-ancilla interaction as well as the Hamiltonian of the 
isolated system itself. Due to the Lieb-Robinson bound, excitations can travel through a locally interacting object 
only with a finite (perhaps averaged in case of a non-homogenous object) velocity $v_I$ (see 
Fig.~\ref{fig Lieb Robinson}). In particular, for an excitation to travel through an object of diameter $d$, it 
takes a time 
\begin{equation}\label{eq time LR}
 \tau = \frac{d}{v_I}. 
\end{equation} 

Like the quantum speed limit, the Lieb-Robinson bound also has a rich history and plays an integral role as a toolkit 
in modern quantum science.\cite{NachtergaeleSimsArXiv2010}

\section{Projective measurement times of finite-size systems}
\label{sec time}

We now have everything we need in order to show that projective measurements on finite-size systems take a finite 
amount of time. To ensure that our arguments are fundamentally valid for all physical systems (irrespective of size), 
we do not impose any restrictions on the degrees of freedom of the measurement apparatus. In particular, we allow 
that some technologically advanced ``alien civilization'' is in possession of a fully controllable, 
fault-tolerant, large scale quantum computer allowing arbitrary pre- and post-processing of arbitrarily many ancillae. 
However, what has to be respected is that the unitary of Eq.~\eqref{eq system ancilla unitary} is implemented by a 
locally-interacting Hamiltonian. 

Two strategies exist for measuring the nonlocal observable, $R_S$, of an object with diameter $d$; either we perform a 
single local measurement by coupling only one ancilla to some part of the system, or we perform many local measurements 
using several ancillae coupled to the macroscopic system at different locations.  

For the first strategy it is sufficient to consider the one-dimensional situation depicted in Fig.~\ref{fig Ising}. 
Computing the quantum speed limit, given in Eq.~\eqref{eq QSL}, now gives us a minimum 
time that varies inversely proportional to the system-ancilla coupling strength, $g_A$.  If the 
interaction is electrostatic, then $g_A=u_\text{Coulomb}/\hbar$ with the Coulomb potential 
\begin{equation}
 u_\text{Coulomb} = k_e\frac{q_S q_A}{r_{SA}}.
\label{eq Coulomb}
\end{equation}
Here, $k_e \approx 9\cdot 10^9$~Nm$^2$/C$^2$ is the Coulomb constant and $r_{SA}$ is the distance between the system 
with charge $q_S$ and the ancilla with charge $q_A$. Our simple model specific result, 
Eq.~(\ref{eq QSL example}), suggests that the quantum speed limit scales like $\tau_\text{sql} \approx \pi/g_A$. 
Equating this with the minimum measurement time, $t_\text{min} = \tau_\text{sql}$, we can write the bound more 
conveniently as 
\begin{equation}
 t_\text{min} = \frac{r_{SA}}{v_E} \qquad \mbox{with} \qquad v_E = \frac{\hbar\pi }{k_e q_S q_A}.
\label{eq tmin}
\end{equation}
Above, we have added the subscript $E$ to emphasize that this speed comes from \emph{energetic} considerations. 

Assuming a single elementary charge, $q_S = q_A = 1.6\cdot10^{-19}$~C (e.g., the ancilla is a single electron, which we let precisely interact with one electron of the system), Eq.~\eqref{eq Coulomb} gives 
\begin{equation}\label{eq QSL Coulomb}
 t_\text{min} = \frac{r_{SA}}{v_E} \qquad \mbox{with} \qquad v_E \approx 10^{5}~\frac{\text{m}}{\text{s}}.
\end{equation}
If the distance between the system and apparatus is $r_{SA} =10^{-10}$~m---of the order of the
size of a hydrogen atom---then we obtain a minimum measurement time of 1 fs. 
Remarkably, this order of magnitude fits surprisingly well cutting-edge technological standards based on 
ultrashort laser pulses, used in femto-chemistry, to study single chemical reactions in real 
time.~\cite{ZewailNobelLec2000} Higher time resolutions are possible using, e.g., free electron lasers, but they 
destroy the system such that the post-measurement state  can no longer be described by 
Eq.~\eqref{eq collapse}.

Let us now consider a macroscopic object of diameter $d=1$~m with an ancilla attached to one end of it. Putting 
$r_{SA}=1$~m in Eq.~\eqref{eq QSL Coulomb} results in a minimum measurement time of $t_\text{min}\approx10^{-5}$~s and corresponds to an ability to resolve processes in the low-frequency radio wave regime, much slower than the timescale many quantum mechanical processes evolve at (e.g., the time scale of chemical reactions in a cat). In turn, this also 
gives us an optimistic estimate of the Lieb-Robinson velocity, $v_I \approx 100,000$~m/s: one order of magnitude larger than the speed of sound in very stiff materials such as diamond. 
Finally, note that for a macroscopic object the distance $r_{SA}$ between the ancilla and the farthest end of the object roughly equals the diameter of the object, i.e.,  $d\approx r_{SA}$. Thus, Eq.~\eqref{eq QSL Coulomb} directly implies 
our central result~\eqref{eq Strasi Skoti Modi time}. 

Can we improve the situation by coupling several ancillae at different locations on the macroscopic system? 
Afterwards, one could imagine that clever post-processing of the ancillae gives rise to a faster measurement. 
For one- and two-dimensional systems this strategy may in fact be possible since one could couple ancillae to every 
part of this object. However, the world around us is three-dimensional and three-dimensional systems are typically 
dominated by their bulk or volume properties. If it is only possible to couple ancillae to the surface of the object, 
then the Lieb-Robinson velocity sets a minimum time needed for some information about the interior of the object to 
influence the ancilla. Taking our above estimate for the Lieb-Robinson velocity, the minimum time based on 
Eq.~\eqref{eq time LR} is 
\begin{equation}
 t_\text{min} = \frac{d}{v_I} \qquad \mbox{with} \qquad v_I \approx 10^{5}~\frac{\text{m}}{\text{s}}.
\end{equation}
In contrast to Eq.~(\ref{eq QSL Coulomb}), the minimum measurement time is here bounded by the speed with which 
\emph{information} can travel through the system. Even though this estimate is independent of the quantum 
speed limit, both estimates agree and are in unison with Eq.~\eqref{eq Strasi Skoti Modi time}. 

While we have focused on electromagnetic interactions, our results are general. Even if we could use strong nuclear 
forces for the system-ancilla interaction, which allow for the measurement of a small object on a yoctosecond 
time-scale ($10^{-24}$~s), the measurement time for a macroscopic object would still be determined by the 
Lieb-Robinson velocity since the strength of the nuclear forces decays very quickly to zero with increasing distance. 
Thus, our bound on the time it takes to measure a \emph{macroscopic} quantum object remains unaffected even if 
it turns out that a quantum measurement of a single microscopic system can be implemented extremely fast. 
Furthermore, even if we consider the speed of light for the Lieb-Robinson velocity, i.e., $v_I = c$, the time 
needed for an excitation to travel a distance of $d = 1$~m would be $\frac{1}{3}10^{-8}$~s. This implies an ability 
to resolve instantaneously processes involving energy differences of 300~MHz (in the low microwave regime). Once more, 
it is impossible to justify applying an instantaneous projective measurement postulate at this time-scale. 

\begin{figure}
 \centering\includegraphics[width=0.40\textwidth,clip=true]{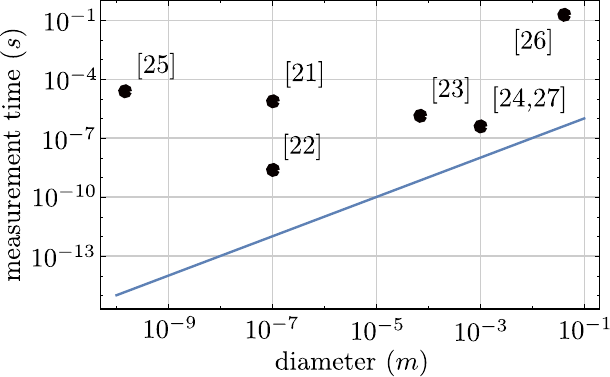}
 \caption{Log-log-plot of the measurement times over the diameter of the measured object. The black dots are 
 experimental data (with the attached number indicating the reference in the bibliography), the blue line is our 
 lower bound. }
 \label{fig experiments} 
\end{figure}

In Fig.~\ref{fig experiments} we compare our main result, the minimum measurement time given in 
Eq.~\eqref{eq Strasi Skoti Modi time} (blue line in Fig.~\ref{fig experiments}), with various state-of-the-art 
experimental platforms~\cite{ElzermanEtAlNat2004, NakajimaEtAlNNT2019, SchaeferEtAlNC2014, RisteEtAlPRL2012, 
HumeRosenbandWinelandPRL2007, NicholsonPRL2012, MonroeEtAlPRL2021} (black dots with reference number in 
Fig.~\ref{fig experiments}; details are presented in Appendix~\ref{sec appendix}). We emphasize that the derivation 
of our bound is based on assuming that macroscopic systems are composed of many locally interacting particles, so we 
do not necessarily expect our bound to hold for a single electron or atom. Nevertheless, as we see in 
Fig.~\ref{fig experiments}, our bound is never violated. 

\section{Discussion}
\label{sec discussion}

We have illustrated how basic physical concepts (in particular, the locality of interactions), which are absent from 
the axioms of quantum theory, can be fruitfully used to learn more about those very axioms. In particular, 
we found that a basic postulate of quantum theory, the instantaneous implementation of a projective 
measurement, is not applicable to macroscopic objects as soon as one takes into account physical considerations. 
Based on intuitive physical reasoning, we derived a lower bound, Eq.~\eqref{eq Strasi Skoti Modi time}, in agreement 
with experimental reality. By shedding some light on the abstract measurement postulate from a physical 
perspective, we hope it appears less confusing for students. At least we showed that, in trying to make 
sense of the measurement postulate, it is possible to introduce many modern and multi-faceted techniques 
of quantum theory. 

The present insights might have also consequences on the \emph{philosophical} questions raised by the ingenious 
thought experiments of Schr\"odinger,\cite{Schroedinger1935a, Schroedinger1935b} Wigner,\cite{WignerInBook1961} and 
others.\cite{DeutschIJTP1985, HardyPRL1992, HardyPRL1993, BruknerInBook2017, FrauchigerRennerNC2018} This is because 
in the mathematical description of these thought experiments cats, or human observers, are often modelled as two-level 
systems which may be subjected to instantaneous measurements in any basis. Whilst these assumptions are compatible 
\emph{in principle} with quantum theory, our analysis shows that they are incompatible with our actual physical world. 

It further remains an open question how tight our bound actually is. However, our bound is likely to be rather 
generous and real macroscopic quantum measurements are likely to take even longer. This is because we only took 
into consideration the time required for the information to be transferred from the system to the ancillae, and do 
not account for the time taken to process the information in the ancillae. 


Finally, our results are of importance for the ongoing endeavor to realize macroscopic quantum states of 
matter in practice. Even if we can create such states the important question remains whether, and how, we can 
\emph{detect} them. In this context it is worth pointing out the related 
studies,\cite{SkotiniotisDuerSekatskiQuantum2017, FroewisEtAlRMP2018, LopezInceraSekatskiDuerQuantum2019} which raised 
similar doubts from a somewhat different perspective than ours. Finally, it is interesting to note that our results 
complement those of Ref.~\cite{GuryanovaFriisHuberQuantum2020} demonstrating that ideal projective measurements require 
infinite resources: if infinitely strong interactions are possible, the lower bound set by the quantum speed limit 
predicts $t_\text{min} = 0$.
 
\emph{Acknowlegdements.} 
This research has received financial support from the DFG (project STR 1505/2-1), the Spanish MINECO FIS2016-80681-P 
(AEI-FEDER, UE), Spanish MICINN PCI2019-111869-2, the Spanish Agencia Estatal de Investigación, project PID2019-107609GB-I00, the Generalitat de Catalunya CIRIT 2017-SGR-1127, the Secretaria d’Universitats i Recerca del Departament d’Empresa i Coneixement de la Generalitat de Catalunya, project ref. 001-P-001644(QuantumCat), the QuantERA grant C’MON-QSENS!, and the Australian Research Council Future Fellowship FT160100073.

\bibliography{/home/philipp/Documents/references/books,/home/philipp/Documents/references/open_systems,/home/philipp/Documents/references/thermo,/home/philipp/Documents/references/info_thermo,/home/philipp/Documents/references/general_QM,/home/philipp/Documents/references/math_phys,/home/philipp/Documents/references/equilibration}

\begin{thebibliography}{41}%
\makeatletter
\providecommand \@ifxundefined [1]{%
 \@ifx{#1\undefined}
}%
\providecommand \@ifnum [1]{%
 \ifnum #1\expandafter \@firstoftwo
 \else \expandafter \@secondoftwo
 \fi
}%
\providecommand \@ifx [1]{%
 \ifx #1\expandafter \@firstoftwo
 \else \expandafter \@secondoftwo
 \fi
}%
\providecommand \natexlab [1]{#1}%
\providecommand \enquote  [1]{``#1''}%
\providecommand \bibnamefont  [1]{#1}%
\providecommand \bibfnamefont [1]{#1}%
\providecommand \citenamefont [1]{#1}%
\providecommand \href@noop [0]{\@secondoftwo}%
\providecommand \href [0]{\begingroup \@sanitize@url \@href}%
\providecommand \@href[1]{\@@startlink{#1}\@@href}%
\providecommand \@@href[1]{\endgroup#1\@@endlink}%
\providecommand \@sanitize@url [0]{\catcode `\\12\catcode `\$12\catcode
  `\&12\catcode `\#12\catcode `\^12\catcode `\_12\catcode `\%12\relax}%
\providecommand \@@startlink[1]{}%
\providecommand \@@endlink[0]{}%
\providecommand \url  [0]{\begingroup\@sanitize@url \@url }%
\providecommand \@url [1]{\endgroup\@href {#1}{\urlprefix }}%
\providecommand \urlprefix  [0]{URL }%
\providecommand \Eprint [0]{\href }%
\providecommand \doibase [0]{http://dx.doi.org/}%
\providecommand \selectlanguage [0]{\@gobble}%
\providecommand \bibinfo  [0]{\@secondoftwo}%
\providecommand \bibfield  [0]{\@secondoftwo}%
\providecommand \translation [1]{[#1]}%
\providecommand \BibitemOpen [0]{}%
\providecommand \bibitemStop [0]{}%
\providecommand \bibitemNoStop [0]{.\EOS\space}%
\providecommand \EOS [0]{\spacefactor3000\relax}%
\providecommand \BibitemShut  [1]{\csname bibitem#1\endcsname}%
\let\auto@bib@innerbib\@empty
\bibitem [{\citenamefont {Wigner}(1963)}]{WignerAJP1963}%
  \BibitemOpen
  \bibfield  {author} {\bibinfo {author} {\bibfnamefont {E.~P.}\ \bibnamefont
  {Wigner}},\ }\bibfield  {title} {\enquote {\bibinfo {title} {{The Problem of
  Measurement}},}\ }\href {https://aapt.scitation.org/doi/10.1119/1.1969254}
  {\bibfield  {journal} {\bibinfo  {journal} {Am. J. Phys.}\ }\textbf {\bibinfo
  {volume} {31}},\ \bibinfo {pages} {6} (\bibinfo {year} {1963})}\BibitemShut
  {NoStop}%
\bibitem [{\citenamefont {Bell}(1990)}]{BellPW1990}%
  \BibitemOpen
  \bibfield  {author} {\bibinfo {author} {\bibfnamefont {J.}~\bibnamefont
  {Bell}},\ }\bibfield  {title} {\enquote {\bibinfo {title} {Against
  'measurement'},}\ }\href
  {https://iopscience.iop.org/article/10.1088/2058-7058/3/8/26/meta?casa_token=XX6A49faji4AAAAA:AklQTb9TIZ0TfHYaiNVon1CC035zh6fQpfLhfeEgBFGRVVTNuB9Qsh4vqBjiPhAGklPJj0fHPvQ}
  {\bibfield  {journal} {\bibinfo  {journal} {Phys. World}\ }\textbf {\bibinfo
  {volume} {3}},\ \bibinfo {pages} {33} (\bibinfo {year} {1990})}\BibitemShut
  {NoStop}%
\bibitem [{\citenamefont {Pokorny}\ \emph {et~al.}(2020)\citenamefont
  {Pokorny}, \citenamefont {Zhang}, \citenamefont {Higgins}, \citenamefont
  {Cabello}, \citenamefont {Kleinmann},\ and\ \citenamefont
  {Hennrich}}]{PokornyEtAlPRL2020}%
  \BibitemOpen
  \bibfield  {author} {\bibinfo {author} {\bibfnamefont {F.}~\bibnamefont
  {Pokorny}}, \bibinfo {author} {\bibfnamefont {C.}~\bibnamefont {Zhang}},
  \bibinfo {author} {\bibfnamefont {G.}~\bibnamefont {Higgins}}, \bibinfo
  {author} {\bibfnamefont {A.}~\bibnamefont {Cabello}}, \bibinfo {author}
  {\bibfnamefont {M.}~\bibnamefont {Kleinmann}}, \ and\ \bibinfo {author}
  {\bibfnamefont {M.}~\bibnamefont {Hennrich}},\ }\bibfield  {title} {\enquote
  {\bibinfo {title} {{Tracking the Dynamics of an Ideal Quantum
  Measurement}},}\ }\href {\doibase 10.1103/PhysRevLett.124.080401} {\bibfield
  {journal} {\bibinfo  {journal} {Phys. Rev. Lett.}\ }\textbf {\bibinfo
  {volume} {124}},\ \bibinfo {pages} {080401} (\bibinfo {year}
  {2020})}\BibitemShut {NoStop}%
\bibitem [{\citenamefont {Cabello}(2017)}]{CabelloInBook2017}%
  \BibitemOpen
  \bibfield  {author} {\bibinfo {author} {\bibfnamefont {A.}~\bibnamefont
  {Cabello}},\ }\enquote {\bibinfo {title} {{What is Quantum Information?}}}\ \
  (\bibinfo  {publisher} {Cambirdge University Press},\ \bibinfo {address}
  {Cambridge},\ \bibinfo {year} {2017})\ Chap.\ \bibinfo {chapter}
  {{Interpretations of Quantum Theory: A Map of Madness}}, pp.\ \bibinfo
  {pages} {138--144}\BibitemShut {NoStop}%
\bibitem [{\citenamefont {Everett}(1957)}]{EverettRMP1957}%
  \BibitemOpen
  \bibfield  {author} {\bibinfo {author} {\bibfnamefont {H.}~\bibnamefont
  {Everett}},\ }\bibfield  {title} {\enquote {\bibinfo {title} {{"Relative
  State" Formulation of Quantum Mechanics}},}\ }\href {\doibase
  10.1103/RevModPhys.29.454} {\bibfield  {journal} {\bibinfo  {journal} {Rev.
  Mod. Phys.}\ }\textbf {\bibinfo {volume} {29}},\ \bibinfo {pages} {454--462}
  (\bibinfo {year} {1957})}\BibitemShut {NoStop}%
\bibitem [{\citenamefont {Holevo}(2001)}]{HolevoBook2001b}%
  \BibitemOpen
  \bibfield  {author} {\bibinfo {author} {\bibfnamefont {A.~S.}\ \bibnamefont
  {Holevo}},\ }\href {\doibase 10.1007/3-540-44998-1} {\emph {\bibinfo {title}
  {Statistical Structure of Quantum Theory}}}\ (\bibinfo  {publisher}
  {Springer-Verlag},\ \bibinfo {address} {Berlin Heidelberg},\ \bibinfo {year}
  {2001})\BibitemShut {NoStop}%
\bibitem [{\citenamefont {D'Alessandro}(2007)}]{DAlessandroBook2007}%
  \BibitemOpen
  \bibfield  {author} {\bibinfo {author} {\bibfnamefont {D.}~\bibnamefont
  {D'Alessandro}},\ }\href@noop {} {\emph {\bibinfo {title} {{Introduction to
  Quantum Control and Dynamics}}}}\ (\bibinfo  {publisher} {Chapman \&
  Hall/CRC},\ \bibinfo {address} {London},\ \bibinfo {year} {2007})\BibitemShut
  {NoStop}%
\bibitem [{\citenamefont {Wiseman}\ and\ \citenamefont
  {Milburn}(2010)}]{WisemanMilburnBook2010}%
  \BibitemOpen
  \bibfield  {author} {\bibinfo {author} {\bibfnamefont {H.~M.}\ \bibnamefont
  {Wiseman}}\ and\ \bibinfo {author} {\bibfnamefont {G.~J.}\ \bibnamefont
  {Milburn}},\ }\href {\doibase 10.1017/CBO9780511813948} {\emph {\bibinfo
  {title} {Quantum Measurement and Control}}}\ (\bibinfo  {publisher}
  {Cambridge University Press},\ \bibinfo {address} {Cambridge},\ \bibinfo
  {year} {2010})\BibitemShut {NoStop}%
\bibitem [{\citenamefont {Jacobs}(2014)}]{JacobsBook2014}%
  \BibitemOpen
  \bibfield  {author} {\bibinfo {author} {\bibfnamefont {K.}~\bibnamefont
  {Jacobs}},\ }\href {\doibase 10.1017/CBO9781139179027} {\emph {\bibinfo
  {title} {Quantum Measurement Theory and its Applications}}}\ (\bibinfo
  {publisher} {Cambridge University Press},\ \bibinfo {address} {Cambridge},\
  \bibinfo {year} {2014})\BibitemShut {NoStop}%
\bibitem [{\citenamefont {Ballentine}(1998)}]{BallentineBook1998}%
  \BibitemOpen
  \bibfield  {author} {\bibinfo {author} {\bibfnamefont {L.~E.}\ \bibnamefont
  {Ballentine}},\ }\href@noop {} {\emph {\bibinfo {title} {{Quantum Mechanics:
  A Modern Development}}}}\ (\bibinfo  {publisher} {World Scientific},\
  \bibinfo {address} {Singapore},\ \bibinfo {year} {1998})\BibitemShut
  {NoStop}%
\bibitem [{\citenamefont {Mandelstam}\ and\ \citenamefont
  {Tamm}(1945)}]{MandelstamTammJP1945}%
  \BibitemOpen
  \bibfield  {author} {\bibinfo {author} {\bibfnamefont {L.}~\bibnamefont
  {Mandelstam}}\ and\ \bibinfo {author} {\bibfnamefont {I.~G.}\ \bibnamefont
  {Tamm}},\ }\bibfield  {title} {\enquote {\bibinfo {title} {The uncertainty
  relation between energy and time in nonrelativistic quantum mechanics},}\
  }\href {https://link.springer.com/chapter/10.1007%2F978-3-642-74626-0_8}
  {\bibfield  {journal} {\bibinfo  {journal} {J. Phys.}\ }\textbf {\bibinfo
  {volume} {9}},\ \bibinfo {pages} {249} (\bibinfo {year} {1945})}\BibitemShut
  {NoStop}%
\bibitem [{\citenamefont {Margolus}\ and\ \citenamefont
  {Levitin}(1998)}]{MargolusLevitinPD1998}%
  \BibitemOpen
  \bibfield  {author} {\bibinfo {author} {\bibfnamefont {N.}~\bibnamefont
  {Margolus}}\ and\ \bibinfo {author} {\bibfnamefont {L.~B.}\ \bibnamefont
  {Levitin}},\ }\bibfield  {title} {\enquote {\bibinfo {title} {The maximum
  speed of dynamical evolution},}\ }\href
  {https://www.sciencedirect.com/science/article/abs/pii/S0167278998000542?via%3Dihub}
  {\bibfield  {journal} {\bibinfo  {journal} {Physica D}\ }\textbf {\bibinfo
  {volume} {120}},\ \bibinfo {pages} {188--195} (\bibinfo {year}
  {1998})}\BibitemShut {NoStop}%
\bibitem [{\citenamefont {Deffner}\ and\ \citenamefont
  {Campbell}(2017)}]{DeffnerCampbellJPA2017}%
  \BibitemOpen
  \bibfield  {author} {\bibinfo {author} {\bibfnamefont {S.}~\bibnamefont
  {Deffner}}\ and\ \bibinfo {author} {\bibfnamefont {S.}~\bibnamefont
  {Campbell}},\ }\bibfield  {title} {\enquote {\bibinfo {title} {Quantum speed
  limits: from {H}eisenberg’s uncertainty principle to optimal quantum
  control},}\ }\href {http://stacks.iop.org/1751-8121/50/i=45/a=453001}
  {\bibfield  {journal} {\bibinfo  {journal} {J. Phys. A}\ }\textbf {\bibinfo
  {volume} {50}},\ \bibinfo {pages} {453001} (\bibinfo {year}
  {2017})}\BibitemShut {NoStop}%
\bibitem [{\citenamefont {Fulde}(1995)}]{FuldeBook1995}%
  \BibitemOpen
  \bibfield  {author} {\bibinfo {author} {\bibfnamefont {P.}~\bibnamefont
  {Fulde}},\ }\href@noop {} {\emph {\bibinfo {title} {Electron Correlations in
  Molecules and Solids}}}\ (\bibinfo  {publisher} {3rd ed., Springer-Verlag},\
  \bibinfo {address} {Berlin},\ \bibinfo {year} {1995})\BibitemShut {NoStop}%
\bibitem [{\citenamefont {Bruus}\ and\ \citenamefont
  {Flensberg}(2004)}]{BruusFlensbergBook2004}%
  \BibitemOpen
  \bibfield  {author} {\bibinfo {author} {\bibfnamefont {H.}~\bibnamefont
  {Bruus}}\ and\ \bibinfo {author} {\bibfnamefont {K.}~\bibnamefont
  {Flensberg}},\ }\href@noop {} {\emph {\bibinfo {title} {Many-body quantum
  theory in condensed matter physics: an introduction}}}\ (\bibinfo
  {publisher} {Oxford University Press},\ \bibinfo {address} {Oxford},\
  \bibinfo {year} {2004})\BibitemShut {NoStop}%
\bibitem [{\citenamefont {Sachdev}(2007)}]{SachdevBook2007}%
  \BibitemOpen
  \bibfield  {author} {\bibinfo {author} {\bibfnamefont {S.}~\bibnamefont
  {Sachdev}},\ }\href@noop {} {\emph {\bibinfo {title} {Quantum Phase
  Transistions}}}\ (\bibinfo  {publisher} {John Wiley \& Sons},\ \bibinfo
  {year} {2007})\BibitemShut {NoStop}%
\bibitem [{\citenamefont {Jaeger}(2014)}]{JaegerAJP2014}%
  \BibitemOpen
  \bibfield  {author} {\bibinfo {author} {\bibfnamefont {G.}~\bibnamefont
  {Jaeger}},\ }\bibfield  {title} {\enquote {\bibinfo {title} {What in the
  (quantum) world is macroscopic?}}\ }\href
  {https://aapt.scitation.org/doi/full/10.1119/1.4878358} {\bibfield  {journal}
  {\bibinfo  {journal} {Am. J. Phys.}\ }\textbf {\bibinfo {volume} {82}},\
  \bibinfo {pages} {896} (\bibinfo {year} {2014})}\BibitemShut {NoStop}%
\bibitem [{\citenamefont {Lieb}\ and\ \citenamefont
  {Robinson}(1972)}]{LiebRobinsonCMP1972}%
  \BibitemOpen
  \bibfield  {author} {\bibinfo {author} {\bibfnamefont {E.~H.}\ \bibnamefont
  {Lieb}}\ and\ \bibinfo {author} {\bibfnamefont {D.~W.}\ \bibnamefont
  {Robinson}},\ }\bibfield  {title} {\enquote {\bibinfo {title} {{The Finite
  Group Velocity of Quantum Spin Systems}},}\ }\href
  {https://link.springer.com/chapter/10.1007/978-3-662-10018-9_25} {\bibfield
  {journal} {\bibinfo  {journal} {Commun. Math. Phys.}\ }\textbf {\bibinfo
  {volume} {28}},\ \bibinfo {pages} {251--257} (\bibinfo {year}
  {1972})}\BibitemShut {NoStop}%
\bibitem [{\citenamefont {Nachtergaele}\ and\ \citenamefont
  {Sims}(2010)}]{NachtergaeleSimsArXiv2010}%
  \BibitemOpen
  \bibfield  {author} {\bibinfo {author} {\bibfnamefont {B.}~\bibnamefont
  {Nachtergaele}}\ and\ \bibinfo {author} {\bibfnamefont {R.}~\bibnamefont
  {Sims}},\ }\bibfield  {title} {\enquote {\bibinfo {title} {{Lieb-Robinson
  Bounds in Quantum Many-Body Physics}},}\ }\href
  {https://arxiv.org/abs/1004.2086} {\bibfield  {journal} {\bibinfo  {journal}
  {arXiv: 1004.2086}\ } (\bibinfo {year} {2010})}\BibitemShut {NoStop}%
\bibitem [{\citenamefont {Zewail}(2000)}]{ZewailNobelLec2000}%
  \BibitemOpen
  \bibfield  {author} {\bibinfo {author} {\bibfnamefont {A.~H.}\ \bibnamefont
  {Zewail}},\ }\bibfield  {title} {\enquote {\bibinfo {title} {{Femtochemistry:
  Atomic-Scale Dynamics of the Chemical Bond Using Ultrafast Lasers (Nobel
  Lecture)}},}\ }\href {\doibase
  10.1002/1521-3773(20000804)39:15<2586::AID-ANIE2586>3.0.CO;2-O} {\bibfield
  {journal} {\bibinfo  {journal} {Angew. Chem. Int. Ed.}\ }\textbf {\bibinfo
  {volume} {39}},\ \bibinfo {pages} {2586--2631} (\bibinfo {year}
  {2000})}\BibitemShut {NoStop}%
\bibitem [{\citenamefont {Elzerman}\ \emph {et~al.}(2004)\citenamefont
  {Elzerman}, \citenamefont {Hanson}, \citenamefont {van Beveren},
  \citenamefont {Witkamp}, \citenamefont {Vandersypen},\ and\ \citenamefont
  {Kouwenhoven}}]{ElzermanEtAlNat2004}%
  \BibitemOpen
  \bibfield  {author} {\bibinfo {author} {\bibfnamefont {J.~M.}\ \bibnamefont
  {Elzerman}}, \bibinfo {author} {\bibfnamefont {R.}~\bibnamefont {Hanson}},
  \bibinfo {author} {\bibfnamefont {L.~H.~Willems}\ \bibnamefont {van
  Beveren}}, \bibinfo {author} {\bibfnamefont {B.}~\bibnamefont {Witkamp}},
  \bibinfo {author} {\bibfnamefont {L.~M.~K.}\ \bibnamefont {Vandersypen}}, \
  and\ \bibinfo {author} {\bibfnamefont {L.~P.}\ \bibnamefont {Kouwenhoven}},\
  }\bibfield  {title} {\enquote {\bibinfo {title} {Single-shot read-out of an
  individual electron spin in a quantum dot},}\ }\href
  {https://www.nature.com/articles/nature02693} {\bibfield  {journal} {\bibinfo
   {journal} {Nature}\ }\textbf {\bibinfo {volume} {430}},\ \bibinfo {pages}
  {431--435} (\bibinfo {year} {2004})}\BibitemShut {NoStop}%
\bibitem [{\citenamefont {Nakajima}\ \emph {et~al.}(2019)\citenamefont
  {Nakajima}, \citenamefont {Noiri}, \citenamefont {Yoneda}, \citenamefont
  {Delbecq}, \citenamefont {Stano}, \citenamefont {Otsuka}, \citenamefont
  {Takeda}, \citenamefont {Amaha}, \citenamefont {Allison}, \citenamefont
  {Kawasaki}, \citenamefont {Ludwig}, \citenamefont {Wieck}, \citenamefont
  {Loss},\ and\ \citenamefont {Tarucha}}]{NakajimaEtAlNNT2019}%
  \BibitemOpen
  \bibfield  {author} {\bibinfo {author} {\bibfnamefont {T.}~\bibnamefont
  {Nakajima}}, \bibinfo {author} {\bibfnamefont {A.}~\bibnamefont {Noiri}},
  \bibinfo {author} {\bibfnamefont {J.}~\bibnamefont {Yoneda}}, \bibinfo
  {author} {\bibfnamefont {M.~R.}\ \bibnamefont {Delbecq}}, \bibinfo {author}
  {\bibfnamefont {P.}~\bibnamefont {Stano}}, \bibinfo {author} {\bibfnamefont
  {T.}~\bibnamefont {Otsuka}}, \bibinfo {author} {\bibfnamefont
  {K.}~\bibnamefont {Takeda}}, \bibinfo {author} {\bibfnamefont
  {S.}~\bibnamefont {Amaha}}, \bibinfo {author} {\bibfnamefont
  {G.}~\bibnamefont {Allison}}, \bibinfo {author} {\bibfnamefont
  {K.}~\bibnamefont {Kawasaki}}, \bibinfo {author} {\bibfnamefont
  {A.}~\bibnamefont {Ludwig}}, \bibinfo {author} {\bibfnamefont {A.~D.}\
  \bibnamefont {Wieck}}, \bibinfo {author} {\bibfnamefont {D.}~\bibnamefont
  {Loss}}, \ and\ \bibinfo {author} {\bibfnamefont {S.}~\bibnamefont
  {Tarucha}},\ }\bibfield  {title} {\enquote {\bibinfo {title} {Quantum
  non-demolition measurement of an electron spin qubit},}\ }\href
  {https://www.nature.com/articles/s41565-019-0426-x} {\bibfield  {journal}
  {\bibinfo  {journal} {Nat. Nanotech.}\ }\textbf {\bibinfo {volume} {14}},\
  \bibinfo {pages} {555--560} (\bibinfo {year} {2019})}\BibitemShut {NoStop}%
\bibitem [{\citenamefont {Sch\"afer}\ \emph {et~al.}(2014)\citenamefont
  {Sch\"afer}, \citenamefont {Herrera}, \citenamefont {Cherukattil},
  \citenamefont {Lovecchio}, \citenamefont {Cataliotti}, \citenamefont
  {Caruso},\ and\ \citenamefont {Smerzi}}]{SchaeferEtAlNC2014}%
  \BibitemOpen
  \bibfield  {author} {\bibinfo {author} {\bibfnamefont {F.}~\bibnamefont
  {Sch\"afer}}, \bibinfo {author} {\bibfnamefont {I.}~\bibnamefont {Herrera}},
  \bibinfo {author} {\bibfnamefont {S.}~\bibnamefont {Cherukattil}}, \bibinfo
  {author} {\bibfnamefont {C.}~\bibnamefont {Lovecchio}}, \bibinfo {author}
  {\bibfnamefont {F.S.}\ \bibnamefont {Cataliotti}}, \bibinfo {author}
  {\bibfnamefont {F.}~\bibnamefont {Caruso}}, \ and\ \bibinfo {author}
  {\bibfnamefont {A.}~\bibnamefont {Smerzi}},\ }\bibfield  {title} {\enquote
  {\bibinfo {title} {Experimental realization of quantum zeno dynamics},}\
  }\href {https://www.nature.com/articles/ncomms4194} {\bibfield  {journal}
  {\bibinfo  {journal} {Nat. Comm.}\ }\textbf {\bibinfo {volume} {5}},\
  \bibinfo {pages} {3194} (\bibinfo {year} {2014})}\BibitemShut {NoStop}%
\bibitem [{\citenamefont {Rist\`e}\ \emph {et~al.}(2012)\citenamefont
  {Rist\`e}, \citenamefont {Bultink}, \citenamefont {Lehnert},\ and\
  \citenamefont {DiCarlo}}]{RisteEtAlPRL2012}%
  \BibitemOpen
  \bibfield  {author} {\bibinfo {author} {\bibfnamefont {D.}~\bibnamefont
  {Rist\`e}}, \bibinfo {author} {\bibfnamefont {C.~C.}\ \bibnamefont
  {Bultink}}, \bibinfo {author} {\bibfnamefont {K.~W.}\ \bibnamefont
  {Lehnert}}, \ and\ \bibinfo {author} {\bibfnamefont {L.}~\bibnamefont
  {DiCarlo}},\ }\bibfield  {title} {\enquote {\bibinfo {title} {{Feedback
  Control of a Solid-State Qubit Using High-Fidelity Projective
  Measurement}},}\ }\href {\doibase 10.1103/PhysRevLett.109.240502} {\bibfield
  {journal} {\bibinfo  {journal} {Phys. Rev. Lett.}\ }\textbf {\bibinfo
  {volume} {109}},\ \bibinfo {pages} {240502} (\bibinfo {year}
  {2012})}\BibitemShut {NoStop}%
\bibitem [{\citenamefont {Hume}\ \emph {et~al.}(2007)\citenamefont {Hume},
  \citenamefont {Rosenband},\ and\ \citenamefont
  {Wineland}}]{HumeRosenbandWinelandPRL2007}%
  \BibitemOpen
  \bibfield  {author} {\bibinfo {author} {\bibfnamefont {D.~B.}\ \bibnamefont
  {Hume}}, \bibinfo {author} {\bibfnamefont {T.}~\bibnamefont {Rosenband}}, \
  and\ \bibinfo {author} {\bibfnamefont {D.~J.}\ \bibnamefont {Wineland}},\
  }\bibfield  {title} {\enquote {\bibinfo {title} {{High-Fidelity Adaptive
  Qubit Detection through Repetitive Quantum Nondemolition Measurements}},}\
  }\href {\doibase 10.1103/PhysRevLett.99.120502} {\bibfield  {journal}
  {\bibinfo  {journal} {Phys. Rev. Lett.}\ }\textbf {\bibinfo {volume} {99}},\
  \bibinfo {pages} {120502} (\bibinfo {year} {2007})}\BibitemShut {NoStop}%
\bibitem [{\citenamefont {Nicholson}\ \emph {et~al.}(2012)\citenamefont
  {Nicholson}, \citenamefont {Martin}, \citenamefont {Williams}, \citenamefont
  {Bloom}, \citenamefont {Bishof}, \citenamefont {Swallows}, \citenamefont
  {Campbell},\ and\ \citenamefont {Ye}}]{NicholsonPRL2012}%
  \BibitemOpen
  \bibfield  {author} {\bibinfo {author} {\bibfnamefont {T.~L.}\ \bibnamefont
  {Nicholson}}, \bibinfo {author} {\bibfnamefont {M.~J.}\ \bibnamefont
  {Martin}}, \bibinfo {author} {\bibfnamefont {J.~R.}\ \bibnamefont
  {Williams}}, \bibinfo {author} {\bibfnamefont {B.~J.}\ \bibnamefont {Bloom}},
  \bibinfo {author} {\bibfnamefont {M.}~\bibnamefont {Bishof}}, \bibinfo
  {author} {\bibfnamefont {M.~D.}\ \bibnamefont {Swallows}}, \bibinfo {author}
  {\bibfnamefont {S.~L.}\ \bibnamefont {Campbell}}, \ and\ \bibinfo {author}
  {\bibfnamefont {J.}~\bibnamefont {Ye}},\ }\bibfield  {title} {\enquote
  {\bibinfo {title} {{Comparison of Two Independent Sr Optical Clocks with
  $1\mathbf{\ifmmode\times\else\texttimes\fi{}}{10}^{\ensuremath{-}17}$
  Stability at ${10}^{3}\text{ }\text{ }\mathbf{s}$}},}\ }\href {\doibase
  10.1103/PhysRevLett.109.230801} {\bibfield  {journal} {\bibinfo  {journal}
  {Phys. Rev. Lett.}\ }\textbf {\bibinfo {volume} {109}},\ \bibinfo {pages}
  {230801} (\bibinfo {year} {2012})}\BibitemShut {NoStop}%
\bibitem [{\citenamefont {Monroe}\ \emph {et~al.}(2021)\citenamefont {Monroe},
  \citenamefont {Yunger~Halpern}, \citenamefont {Lee},\ and\ \citenamefont
  {Murch}}]{MonroeEtAlPRL2021}%
  \BibitemOpen
  \bibfield  {author} {\bibinfo {author} {\bibfnamefont {J.~T.}\ \bibnamefont
  {Monroe}}, \bibinfo {author} {\bibfnamefont {N.}~\bibnamefont
  {Yunger~Halpern}}, \bibinfo {author} {\bibfnamefont {T.}~\bibnamefont {Lee}},
  \ and\ \bibinfo {author} {\bibfnamefont {K.~W.}\ \bibnamefont {Murch}},\
  }\bibfield  {title} {\enquote {\bibinfo {title} {Weak measurement of a
  superconducting qubit reconciles incompatible operators},}\ }\href {\doibase
  10.1103/PhysRevLett.126.100403} {\bibfield  {journal} {\bibinfo  {journal}
  {Phys. Rev. Lett.}\ }\textbf {\bibinfo {volume} {126}},\ \bibinfo {pages}
  {100403} (\bibinfo {year} {2021})}\BibitemShut {NoStop}%
\bibitem [{\citenamefont
  {Schr\"odinger}(1935{\natexlab{a}})}]{Schroedinger1935a}%
  \BibitemOpen
  \bibfield  {author} {\bibinfo {author} {\bibfnamefont {E.}~\bibnamefont
  {Schr\"odinger}},\ }\bibfield  {title} {\enquote {\bibinfo {title} {{Die
  gegenw\"artige Situation in der Quantenmechanik}},}\ }\href
  {https://link.springer.com/article/10.1007/BF01491891} {\bibfield  {journal}
  {\bibinfo  {journal} {Naturwissenschaften}\ }\textbf {\bibinfo {volume}
  {23}},\ \bibinfo {pages} {807--812} (\bibinfo {year}
  {1935}{\natexlab{a}})}\BibitemShut {NoStop}%
\bibitem [{\citenamefont
  {Schr\"odinger}(1935{\natexlab{b}})}]{Schroedinger1935b}%
  \BibitemOpen
  \bibfield  {author} {\bibinfo {author} {\bibfnamefont {E.}~\bibnamefont
  {Schr\"odinger}},\ }\bibfield  {title} {\enquote {\bibinfo {title} {{Die
  gegenw\"artige Situation in der Quantenmechanik}},}\ }\href
  {https://link.springer.com/article/10.1007%2FBF01491914} {\bibfield
  {journal} {\bibinfo  {journal} {Naturwissenschaften}\ }\textbf {\bibinfo
  {volume} {23}},\ \bibinfo {pages} {823--828} (\bibinfo {year}
  {1935}{\natexlab{b}})}\BibitemShut {NoStop}%
\bibitem [{\citenamefont {Wigner}(1961)}]{WignerInBook1961}%
  \BibitemOpen
  \bibfield  {author} {\bibinfo {author} {\bibfnamefont {E.~P.}\ \bibnamefont
  {Wigner}},\ }\enquote {\bibinfo {title} {{The Scientist Speculates}},}\ \
  (\bibinfo  {publisher} {Heinemann},\ \bibinfo {address} {London},\ \bibinfo
  {year} {1961})\ Chap.\ \bibinfo {chapter} {Remarks on the mind-body
  question}, pp.\ \bibinfo {pages} {284--302}\BibitemShut {NoStop}%
\bibitem [{\citenamefont {Deutsch}(1985)}]{DeutschIJTP1985}%
  \BibitemOpen
  \bibfield  {author} {\bibinfo {author} {\bibfnamefont {D.}~\bibnamefont
  {Deutsch}},\ }\bibfield  {title} {\enquote {\bibinfo {title} {Quantum theory
  as a universal physical theory},}\ }\href {\doibase 10.1007/BF00670071}
  {\bibfield  {journal} {\bibinfo  {journal} {Int. J. Theor. Phys.}\ }\textbf
  {\bibinfo {volume} {24}},\ \bibinfo {pages} {1--41} (\bibinfo {year}
  {1985})}\BibitemShut {NoStop}%
\bibitem [{\citenamefont {Hardy}(1992)}]{HardyPRL1992}%
  \BibitemOpen
  \bibfield  {author} {\bibinfo {author} {\bibfnamefont {L.}~\bibnamefont
  {Hardy}},\ }\bibfield  {title} {\enquote {\bibinfo {title} {{Quantum
  mechanics, local realistic theories, and Lorentz-invariant realistic
  theories}},}\ }\href {\doibase 10.1103/PhysRevLett.68.2981} {\bibfield
  {journal} {\bibinfo  {journal} {Phys. Rev. Lett.}\ }\textbf {\bibinfo
  {volume} {68}},\ \bibinfo {pages} {2981--2984} (\bibinfo {year}
  {1992})}\BibitemShut {NoStop}%
\bibitem [{\citenamefont {Hardy}(1993)}]{HardyPRL1993}%
  \BibitemOpen
  \bibfield  {author} {\bibinfo {author} {\bibfnamefont {L.}~\bibnamefont
  {Hardy}},\ }\bibfield  {title} {\enquote {\bibinfo {title} {Nonlocality for
  two particles without inequalities for almost all entangled states},}\ }\href
  {\doibase 10.1103/PhysRevLett.71.1665} {\bibfield  {journal} {\bibinfo
  {journal} {Phys. Rev. Lett.}\ }\textbf {\bibinfo {volume} {71}},\ \bibinfo
  {pages} {1665--1668} (\bibinfo {year} {1993})}\BibitemShut {NoStop}%
\bibitem [{\citenamefont {Brukner}(2017)}]{BruknerInBook2017}%
  \BibitemOpen
  \bibfield  {author} {\bibinfo {author} {\bibfnamefont {\v{C}.}\ \bibnamefont
  {Brukner}},\ }\enquote {\bibinfo {title} {{Quantum [Un]Speakables II: Half a
  Century of Bell's Theorem}},}\ \ (\bibinfo  {publisher} {Springer},\ \bibinfo
  {year} {2017})\ Chap.\ \bibinfo {chapter} {On the quantum measurement
  problem}, pp.\ \bibinfo {pages} {95--117}\BibitemShut {NoStop}%
\bibitem [{\citenamefont {Frauchiger}\ and\ \citenamefont
  {Renner}(2018)}]{FrauchigerRennerNC2018}%
  \BibitemOpen
  \bibfield  {author} {\bibinfo {author} {\bibfnamefont {D.}~\bibnamefont
  {Frauchiger}}\ and\ \bibinfo {author} {\bibfnamefont {R.}~\bibnamefont
  {Renner}},\ }\bibfield  {title} {\enquote {\bibinfo {title} {Quantum theory
  cannot consistently describe the use of itself},}\ }\href
  {https://www.nature.com/articles/s41467-018-05739-8/} {\bibfield  {journal}
  {\bibinfo  {journal} {Nat. Comm.}\ }\textbf {\bibinfo {volume} {9}},\
  \bibinfo {pages} {3711} (\bibinfo {year} {2018})}\BibitemShut {NoStop}%
\bibitem [{\citenamefont {Skotiniotis}\ \emph {et~al.}(2017)\citenamefont
  {Skotiniotis}, \citenamefont {D{\"{u}}r},\ and\ \citenamefont
  {Sekatski}}]{SkotiniotisDuerSekatskiQuantum2017}%
  \BibitemOpen
  \bibfield  {author} {\bibinfo {author} {\bibfnamefont {M.}~\bibnamefont
  {Skotiniotis}}, \bibinfo {author} {\bibfnamefont {W.}~\bibnamefont
  {D{\"{u}}r}}, \ and\ \bibinfo {author} {\bibfnamefont {P.}~\bibnamefont
  {Sekatski}},\ }\bibfield  {title} {\enquote {\bibinfo {title} {Macroscopic
  superpositions require tremendous measurement devices},}\ }\href {\doibase
  10.22331/q-2017-11-21-34} {\bibfield  {journal} {\bibinfo  {journal}
  {{Quantum}}\ }\textbf {\bibinfo {volume} {1}},\ \bibinfo {pages} {34}
  (\bibinfo {year} {2017})}\BibitemShut {NoStop}%
\bibitem [{\citenamefont {Fr\"owis}\ \emph {et~al.}(2018)\citenamefont
  {Fr\"owis}, \citenamefont {Sekatski}, \citenamefont {D\"ur}, \citenamefont
  {Gisin},\ and\ \citenamefont {Sangouard}}]{FroewisEtAlRMP2018}%
  \BibitemOpen
  \bibfield  {author} {\bibinfo {author} {\bibfnamefont {F.}~\bibnamefont
  {Fr\"owis}}, \bibinfo {author} {\bibfnamefont {P.}~\bibnamefont {Sekatski}},
  \bibinfo {author} {\bibfnamefont {W.}~\bibnamefont {D\"ur}}, \bibinfo
  {author} {\bibfnamefont {N.}~\bibnamefont {Gisin}}, \ and\ \bibinfo {author}
  {\bibfnamefont {N.}~\bibnamefont {Sangouard}},\ }\bibfield  {title} {\enquote
  {\bibinfo {title} {Macroscopic quantum states: Measures, fragility, and
  implementations},}\ }\href {\doibase 10.1103/RevModPhys.90.025004} {\bibfield
   {journal} {\bibinfo  {journal} {Rev. Mod. Phys.}\ }\textbf {\bibinfo
  {volume} {90}},\ \bibinfo {pages} {025004} (\bibinfo {year}
  {2018})}\BibitemShut {NoStop}%
\bibitem [{\citenamefont {L{\'{o}}pez-Incera}\ \emph
  {et~al.}(2019)\citenamefont {L{\'{o}}pez-Incera}, \citenamefont {Sekatski},\
  and\ \citenamefont {D{\"{u}}r}}]{LopezInceraSekatskiDuerQuantum2019}%
  \BibitemOpen
  \bibfield  {author} {\bibinfo {author} {\bibfnamefont {A.}~\bibnamefont
  {L{\'{o}}pez-Incera}}, \bibinfo {author} {\bibfnamefont {P.}~\bibnamefont
  {Sekatski}}, \ and\ \bibinfo {author} {\bibfnamefont {W.}~\bibnamefont
  {D{\"{u}}r}},\ }\bibfield  {title} {\enquote {\bibinfo {title} {All
  macroscopic quantum states are fragile and hard to prepare},}\ }\href
  {\doibase 10.22331/q-2019-01-25-118} {\bibfield  {journal} {\bibinfo
  {journal} {{Quantum}}\ }\textbf {\bibinfo {volume} {3}},\ \bibinfo {pages}
  {118} (\bibinfo {year} {2019})}\BibitemShut {NoStop}%
\bibitem [{\citenamefont {Guryanova}\ \emph {et~al.}(2020)\citenamefont
  {Guryanova}, \citenamefont {Friis},\ and\ \citenamefont
  {Huber}}]{GuryanovaFriisHuberQuantum2020}%
  \BibitemOpen
  \bibfield  {author} {\bibinfo {author} {\bibfnamefont {Y.}~\bibnamefont
  {Guryanova}}, \bibinfo {author} {\bibfnamefont {N.}~\bibnamefont {Friis}}, \
  and\ \bibinfo {author} {\bibfnamefont {M.}~\bibnamefont {Huber}},\ }\bibfield
   {title} {\enquote {\bibinfo {title} {Ideal {P}rojective {M}easurements
  {H}ave {I}nfinite {R}esource {C}osts},}\ }\href {\doibase
  10.22331/q-2020-01-13-222} {\bibfield  {journal} {\bibinfo  {journal}
  {{Quantum}}\ }\textbf {\bibinfo {volume} {4}},\ \bibinfo {pages} {222}
  (\bibinfo {year} {2020})}\BibitemShut {NoStop}%
\bibitem [{\citenamefont {Paik}\ \emph {et~al.}(2011)\citenamefont {Paik},
  \citenamefont {Schuster}, \citenamefont {Bishop}, \citenamefont {Kirchmair},
  \citenamefont {Catelani}, \citenamefont {Sears}, \citenamefont {Johnson},
  \citenamefont {Reagor}, \citenamefont {Frunzio}, \citenamefont {Glazman},
  \citenamefont {Girvin}, \citenamefont {Devoret},\ and\ \citenamefont
  {Schoelkopf}}]{PaikEtAlPRL2011}%
  \BibitemOpen
  \bibfield  {author} {\bibinfo {author} {\bibfnamefont {Hanhee}\ \bibnamefont
  {Paik}}, \bibinfo {author} {\bibfnamefont {D.~I.}\ \bibnamefont {Schuster}},
  \bibinfo {author} {\bibfnamefont {Lev~S.}\ \bibnamefont {Bishop}}, \bibinfo
  {author} {\bibfnamefont {G.}~\bibnamefont {Kirchmair}}, \bibinfo {author}
  {\bibfnamefont {G.}~\bibnamefont {Catelani}}, \bibinfo {author}
  {\bibfnamefont {A.~P.}\ \bibnamefont {Sears}}, \bibinfo {author}
  {\bibfnamefont {B.~R.}\ \bibnamefont {Johnson}}, \bibinfo {author}
  {\bibfnamefont {M.~J.}\ \bibnamefont {Reagor}}, \bibinfo {author}
  {\bibfnamefont {L.}~\bibnamefont {Frunzio}}, \bibinfo {author} {\bibfnamefont
  {L.~I.}\ \bibnamefont {Glazman}}, \bibinfo {author} {\bibfnamefont {S.~M.}\
  \bibnamefont {Girvin}}, \bibinfo {author} {\bibfnamefont {M.~H.}\
  \bibnamefont {Devoret}}, \ and\ \bibinfo {author} {\bibfnamefont {R.~J.}\
  \bibnamefont {Schoelkopf}},\ }\bibfield  {title} {\enquote {\bibinfo {title}
  {{Observation of High Coherence in Josephson Junction Qubits Measured in a
  Three-Dimensional Circuit QED Architecture}},}\ }\href {\doibase
  10.1103/PhysRevLett.107.240501} {\bibfield  {journal} {\bibinfo  {journal}
  {Phys. Rev. Lett.}\ }\textbf {\bibinfo {volume} {107}},\ \bibinfo {pages}
  {240501} (\bibinfo {year} {2011})}\BibitemShut {NoStop}%
\bibitem [{\citenamefont {Naghiloo}(2019)}]{NaghilooPhD}%
  \BibitemOpen
  \bibfield  {author} {\bibinfo {author} {\bibfnamefont {M.}~\bibnamefont
  {Naghiloo}},\ }\emph {\bibinfo {title} {{Introduction to Experimental Quantum
  Measurement with Superconducting Qubits}}},\ \href@noop {} {Ph.D. thesis},\
  \bibinfo  {school} {Washington University} (\bibinfo {year}
  {2019})\BibitemShut {NoStop}%
\end{thebibliography}%

\appendix
\section{The Zassenhaus Formula}\label{sec appendix Zassenhaus}
Let $X, Y$, be two Hermitian operators.  Zassenhaus' formula states
\begin{equation}
 \begin{split}\label{eq Zasse}
  e^{t(X+Y)} =& e^{tX}\, e^{tY}\, e^{-\frac{t^2}{2}[X,Y]}\, \\
  & \times e^{\frac{t^3}{6}(2[Y,[X,Y]]+[X,[X,Y]])}\ldots,
 \end{split}
\end{equation}
where the terms that follow are exponentials of higher-order nested commutators.  To derive Zassenhaus' formula one simply
Taylor expands the exponentials and collects terms.  

To see how one can use Zassenhaus' formula to engineer long-range interactions we return to the one-dimensional Ising 
interaction considered in the main text (Eq.~\eqref{eq system ancilla Ising}).  Setting $X=H_\text{Ising}$ and 
$Y= h_A\sigma^{(z)}_A + g_A\sigma^{(x)}_N\sigma^{(x)}_A$, one sees that if all the high-order nested commutators in 
Eq.~\eqref{eq Zasse} are non-zero, then by controlling $h_A, g_A$ in Eq.~\eqref{eq system ancilla Ising} one can engineer a large class of unitaries.  It remains a challenging and interesting problem in quantum control theory 
whether with such limited ``local'' control one is able to generate highly non-trivial global interactions, but 
it is possible in principle.~\cite{DAlessandroBook2007} 

\section{Experimental data for measurement times}
\label{sec appendix}

Ref.~\cite{ElzermanEtAlNat2004} reports on the readout of the spin state of an electron confined in a quantum dot with a size of roughly 100~nm. The readout timescale is reported to be 8~$\mu$s. 

In Ref.~\cite{NakajimaEtAlNNT2019} a quantum non-demolition measurement of a similar system (a quantum dot) was reported using an ancilla of two adjacent quantum dots. The smallest measurement time $\tau_k$ for $k=1$ was 
reported to be 2.33~ns.

Ref.~\cite{SchaeferEtAlNC2014} reports on experimentally observed quantum Zeno dynamics in a Bose-Einstein condensate of Rubidium atoms. To estimate the size of the atomic cloud, we use the waist of the laser beam (70~$\mu$m). To 
estimate the measurement time, we add the 0.8~$\mu$s for the $\pi$-pulse created by a Raman beam to the 0.6~$\mu$s 
required for the time to illuminate the condensate with dissipative light (in total 1.4~$\mu$s). 

Ref.~\cite{RisteEtAlPRL2012} reports on projective measurements of a superconducting transmon qubit, whose size we 
estimate based on Ref.~\cite{PaikEtAlPRL2011} to be 1~mm, using a microwave pulse of 400~ns, which we set as our 
measurement time. 

Ref.~\cite{HumeRosenbandWinelandPRL2007} reports an ancilla-assisted read-out of a single Aluminium ion. The interaction time between the Aluminium ion and the ancilla (a Beryllium ion) is quoted with $25~\mu$s. The diameter of 
the object is estimated using the radius of an Aluminium atom ($143~$pm). 

Ref.~\cite{NicholsonPRL2012} reports on the read-out of optical lattice clocks using Rabi spectroscopy with a probe time of $160~$ms. To estimate the diameter of the sample, we multiply the lattice constant of $813~$nm with the number 
of atoms, which can be scaled up to 50.000 ``under typical experimental conditions.'' Thus, the diameter of the object is roughly $4~$cm. 

Ref.~\cite{MonroeEtAlPRL2021} reports on projective measurements of a superconducting transmon qubit on a 
time-scale of 350~ns. As for Ref.~\cite{RisteEtAlPRL2012}, the size of the transmon qubit in this experiment is 
estimated to be 1~mm.\cite{NaghilooPhD} Note that the 50~ns faster read-out in Ref.~\cite{MonroeEtAlPRL2021} 
compared to Ref.~\cite{RisteEtAlPRL2012} is invisible in Fig.~\ref{fig experiments} due to the logarithmic scale 
covering many orders of magnitude. 
 
\end{document}